\documentclass{elsart}
\usepackage[dvips]{graphicx}
\newcommand{\wigner}{\ensuremath{d_{00}^\ell (\costh)}}
\newcommand{\ess}{\ensuremath{S(\mkpi)}}

\newcommand{\phidk}{\ensuremath{\theta^*_{K^+ \pi^+}}}
\newcommand{\cosphidk}{\ensuremath{\cos{\phidk}}}
\newcommand{\pee}{\ensuremath{P(\mkpi)}}

\newcommand{\dee}{\ensuremath{D(\mkpi)}}

\newcommand{\SP}{\ensuremath{2~\ess \times \pee}}
\newcommand{\SD}{\ensuremath{2~\ess \times \dee}}
\newcommand{\PD}{\ensuremath{2~\pee \times \dee}}
\newcommand{\Dint}{\ensuremath{D^2~(\mkpi)}}
\newcommand{\Sint}{\ensuremath{S^2~(\mkpi)}}
\newcommand{\Pint}{\ensuremath{P^2~(\mkpi)}}

\newcommand{\kk}{\ensuremath{K^- K^+}}
\newcommand{\kpi}{\ensuremath{K^- \pi^+}}

\newcommand{\krz}{\ensuremath{\overline{K}^{*0}}}
\newcommand{\ksw}{\ensuremath{\overline{K}_0^{*0}(1430)}}
\newcommand{\krzb}{\ensuremath{\overline{K}^{*0}}}
\newcommand{\krzmndk}{\ensuremath{D^+ \rightarrow \krzb e^+ \nu}}

\newcommand{\kpilndk}{\ensuremath{D^+ \rightarrow K^- \pi^+ \ell^+ \nu }}

\newcommand{\costh}{\ensuremath{\cos\theta}}

\newcommand{\qsq}{\ensuremath{q^2}}

\newcommand{\mkpi}{\ensuremath{m_{K^-\pi^+}}}
\newcommand{\mnr}{\ensuremath{m_{K^+\pi^+}}}
\newcommand{\mkk}{\ensuremath{m_{K^+ K^-}}}

\newcommand{\dpkkpi}{\ensuremath{D^+ \rightarrow K^- K^+ \pi^+ }}
\newcommand{\dpkpipi}{\ensuremath{D^+ \rightarrow K^- \pi^+ \pi^+ }}
\newcommand{\dskkpi}{\ensuremath{D_s^+ \rightarrow K^- K^+ \pi^+ }}
\newcommand{\fz}{\ensuremath{f_0(980)}}

\newcommand{\mevc}{\ensuremath{\textrm{MeV}/c^2}}

\newcommand{\mysection}[1]{\section{#1}}
\newcounter{saveeqn}%

\begin{document}
\begin{frontmatter}
\title{A Non-parametric Approach to Measuring the \kpi{} Amplitudes in  \dpkkpi{} Decay }
\collaboration{The~FOCUS~Collaboration}\footnotemark
\author[ucd]{J.~M.~Link}
\author[ucd]{P.~M.~Yager}
\author[cbpf]{J.~C.~Anjos}
\author[cbpf]{I.~Bediaga}
\author[cbpf]{C.~Castromonte}
\author[cbpf]{A.~A.~Machado}
\author[cbpf]{J.~Magnin}
\author[cbpf]{A.~Massafferri}
\author[cbpf]{J.~M.~de~Miranda}
\author[cbpf]{I.~M.~Pepe}
% ...

\address[ucd]{University of California, Davis, CA 95616}
\address[cbpf]{Centro Brasileiro de Pesquisas F\'\i sicas, Rio de Janeiro, RJ, Brazil}
% ...

\footnotetext{See \textrm{http://www-focus.fnal.gov/authors.html} for additional author information.}
\nobreak
\begin{abstract}
Using a large sample of \dpkkpi{} decays
collected by the FOCUS photoproduction experiment at Fermilab, we
present the first non-parametric analysis of the \kpi{} amplitudes in \dpkkpi{} decay. The technique is similar to the technique used for our non-parametric measurements of the \krzmndk{} form factors. Although these results are in rough agreement with those of E687, we observe a wider S-wave contribution for the \ksw{} contribution than the standard, PDG~\cite{pdg} Breit-Wigner parameterization. We have some weaker evidence
for the existence of a new, D-wave component at low values of the $K^- \pi^+$ mass.
\end{abstract}
\end{frontmatter}


\mysection{Introduction}
This paper describes a non-parametric measurement of the \kpi{}
amplitudes present in the decay \dpkkpi{}.  Charm decay Dalitz plots are traditionally fit
using the isobar model where the amplitude is represented by a sum of known Breit-Wigner resonances multiplied
by complex amplitudes along with a possible non-resonant term~\cite{asner}.  
Often this approach gives a good qualitative
representation of the observed populations in the Dalitz plot. The isobar approach, however, does not automatically
incorporate some important theoretical constraints.  If all final state interactions are dominated by the
two-body resonant system, with negligible contribution from the third
body, then two-body unitarity is  violated by the isobar model. 
These potential unitarity violations are particularly severe for the case of broad, overlapping resonances.  

An alternative, K-matrix formalism spearheaded by the FOCUS collaboration \cite{Kmatrix} for charm meson decays, is designed to automatically satisfy unitarity but it is a difficult analysis that 
requires considerable input from low energy scattering experiments on the form and location of K-matrix poles.  In particular the FOCUS analysis of the $D^+,~D_s^+ \rightarrow \pi^+ \pi^- \pi^+ $ final state used the K-matrix
description for just the broad dipion states, but the narrow P-wave states were incorporated as isobar contributions.

This work takes a considerably different approach. We directly measure the \kpi{} spin amplitudes 
as a function of  \mkpi{} mass by ``projecting" them based on the decay angular distribution.  
Our projective weighting technique, described in Section \ref{proj}, is very similar to that used to make non-parametric
measurements \cite{helicity} of the \qsq{} dependence of the helicity amplitudes in \kpilndk{}. In our earlier work, each helicity amplitude is projected by making a weighted histogram of \qsq{} using special weights designed to block all other amplitude contributions. The projective weighting
technique is therefore an intrinsically one dimensional analysis.  
The \dpkkpi{} final state -- in 
principle -- is influenced by three amplitudes: $K^- \pi^+$, $K^- K^+$, and even $K^+ \pi^+$.
We have chosen the \dpkkpi{} final state as a first test case since the E687 isobar analysis \cite{rlg} concluded that the observed \dpkkpi{} Dalitz plot could be adequately described by just 
three resonant contributions: $\phi \pi^+$, $K^+ \krz$, and $K^+ \ksw$.  Although $\phi \pi^+$
is an important contribution, the $\phi$ is a very narrow resonance that can be substantially
removed by placing a lower cut on the \mkk{} mass (such as $\mkk{} > 1050~\mevc$). Because
there is no overlap of the $\phi$ band with the \krzb{} band and most of the kinematically allowed \ksw{} region, there
is a relatively small loss of information from the anti-$\phi$ cut.  The technique
used to correct for residual $\phi$ contamination is described in Section \ref{bias}. It is important to establish to what extent undiscovered contributions in the $K^+ K^-$ channel such as $\fz{} \pi^+$ could influence
these results.\footnote{Evidence for a small $a_0(980)$ contribution based on an analysis of FOCUS data was reported in  
a conference proceeding \cite{cipanp}.} For example, the $\fz{} \pi^+$ contribution was found \cite{rlg} to be a major contributing channel in the related $D_s^+ \rightarrow K^+ K^- \pi^+$ decay.  Uncertainty in the amplitudes describing the \kk{} channel will form the major systematic error in this analysis.
Throughout this paper, unless explicitly stated otherwise,
the charge conjugate state is implied when a decay mode of a specific
charge is stated.

\mysection{\label{exp} Experimental and analysis details}

\begin{figure}[tbph!]
\begin{center}
\includegraphics[width=3.in]{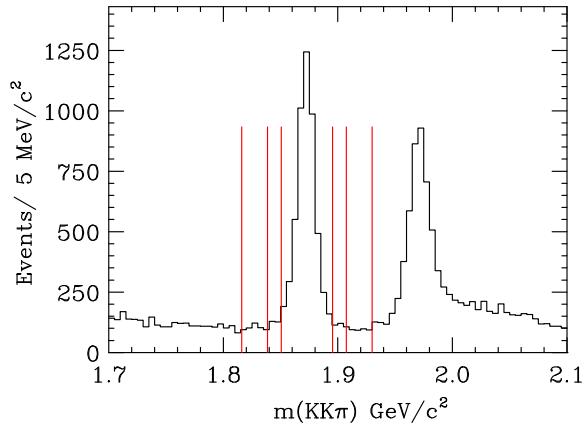}
\end{center}
\caption{ The $K^- K^+ \pi^+$ invariant mass spectra for all of the cuts used
in this analysis. The \dpkkpi{} signal and sideband regions are
shown with vertical lines.  The peak to the right of the \dpkkpi{}peak is due
to \dskkpi{}.  The $D_s^+$ peak appears over a broad, ramped background due to 
misidentified \dpkpipi{} decays.
\label{signal}}
\end{figure}

The data for this paper were collected in the Wideband photoproduction
experiment FOCUS during the Fermilab 1996--1997 fixed-target run. In
FOCUS, a forward multi-particle spectrometer is used to measure the
interactions of high energy photons in a segmented BeO target. The
FOCUS detector is a large aperture, fixed-target spectrometer with
excellent track resolution and particle identification. Most of the FOCUS
experiment and analysis techniques have been described
previously~\cite{nim,CNIM,VNIM,SNIM}.
To obtain the signal displayed in figure \ref{signal}, we required
that the $K^- K^+ \pi^+$ tracks formed a vertex with a confidence level
in excess of 10\%, and the $K^- K^+ \pi^+$  vertex was outside of our 
BeO target and all other spectrometer material by at least 3$\sigma$. Other tracks
along with information from the  
$K^- K^+ \pi^+$  ``secondary" vertex were used for form a ``primary" vertex 
and the separation between the primary and
secondary vertex was required to exceed 7$\sigma$. The kaon hypothesis was favored over the pion
hypothesis by 3 units of log likelihood in our Cerenkov system for the $K^-$ and $K^+$ candidates,
and the $\pi^+$ candidate response was consistent with that for a pion.  Finally, we
required that the secondary vertex was well isolated. We required that no 
primary vertex track was consistent with the secondary vertex with a confidence
level exceeding 0.5 \% , and no other track in the event was consistent with the secondary
vertex with a confidence level exceeding $1 \times 10^{-4}$.  In this analysis
we eliminated most of the $D^+ \rightarrow \phi \pi^+$ contribution by imposing an
anti-$\phi$ cut that required \mkk{} $> ~1050~\mevc$.  

We obtained a signal of 6400 \dpkkpi{} events 
prior to the anti-$\phi$ cut and 4200 events after the anti-$\phi$ cut.

\mysection{\label{proj} Projection Weighting Technique} 
The projector method used in this analysis is nearly identical to that used to determine the \qsq{}~dependence of the helicity form factor in \kpilndk{} decay \cite{helicity}.
To apply the projector method, we must assume that after we impose the anti-$\phi$ cut, the residual effects of the \kk{} amplitude contributions are small enough that
we can reliably correct for them using the bias technique described in Section \ref{bias}.
In the absence of \kk{} resonances, we can write the decay amplitude in terms
of $\mkpi = m$ and the decay angle $\theta$ (which is the angle between the $K^-$ and the $K^+$ 
in the $K^- \pi^+$ rest frame). 
\[
A{\rm{ }} = \sum\limits_\ell ^{s,p,d \cdots } {A_\ell } \left( {m } \right)~\wigner{}  
\]
Here \wigner{} are the Wigner d-matrices that describe the amplitude for a 
\kpi{} system of angular momentum $\ell$ to simultaneously have 0 angular momentum along its 
($D^+$ frame) helicity axis and the \kpi decay axis. 
For simplicity, we illustrate the technique assuming only P-wave and S-wave contributions although contributions up to and including D-wave are included in the analysis. In the absence of D-wave or 
higher amplitudes, the decay intensity consists of three terms which depend on the two complex functions: $S(m)$ and $P(m)$ of $m \equiv \mkpi$.

\[
\begin{array}{l}
 I(m{\rm{ }},\,\cos \theta ) = \left| A \right|^2  = \left| {\;S(m) + P(m)\cos \theta \;} \right|^2  =  \\ 
 \left| {S(m)} \right|^2  + 2{\mathop{\rm Re}\nolimits} \left\{ {S^ *  (m)\;P(m)} \right\}\cos \theta  + \left| {P(m)} \right|^2 \cos ^2 \theta  \\ 
 \end{array}
\]
For notational simplicity we will write the direct terms as 
$SS(m) \equiv \left| {S(m)} \right|^2$ and $PP(m) \equiv \left| {P(m)} \right|^2 $ and the interference
term as $SP(m) \equiv \mathop{\rm Re} \left\{ {S^ *  (m)\;P(m)} \right\}$:
\begin{eqnarray}
I(m{\rm{ }},\cos \theta ) = SS\left( m \right) + 2 {SP\left( m \right)}\cos \theta  + PP(m)\cos^2 \theta 
\label{intensity}
\end{eqnarray}
Our approach is to divide $\cos{\theta}$ into twenty evenly spaced angular bins.
Let ${}^i\vec D = \left( {\begin{array}{*{20}c}
   {{}^in_1 } & {{}^in_2 } & {...} & {{}^in_{20} }  \\
\end{array}} \right)$ be a vector whose 20 components give the population in data for each of the 20 $\cos{\theta}$ bins. Here $^i$ specifies the $i^{th}$ \mkpi{} bin.
Our goal is to represent the  ${}^i\vec D$ vector as a sum over the expected
populations for each of the three partial waves.  We will call these ${}^i \vec m$
vectors. For this simplified case, there are three such vectors computed for each \mkpi{} bin:
$\left\{ {}^i {\vec m_\alpha  } \right\} = 
   \left({{}^i\vec m_{SS} } , {{}^i\vec m_{SP} } , {{}^i\vec m_{PP} } \right)$.  
Each ${}^i \vec m_\alpha$ is generated using a 
phase space and full detector simulation for \dpkkpi{} decay with one amplitude turned on, and all other amplitudes shut off.  Hence the ${}^i \vec m_{PP}$
vector is computed assuming an intensity of $\cos^2{\theta}$ for each \mkpi{}
bin. The ${}^i \vec m$ vectors incorporate
the underlying angular distribution as well as all acceptance and 
cut efficiency effects. In particular, the anti-$\phi$ cut creates substantial inefficiencies at low \costh{} and high \mkpi{}, whereas the other cuts have a reasonably uniform acceptance in \costh{}.

We use a weighting technique to fit the bin populations in the data to the form:
$
{}^i \vec D  = F_{SS} \left( {m_i } \right)\;^i \vec m_{SS}  + F_{SP} \left( {m_i } \right)\;^i \vec m_{SP}  + F_{PP} \left( {m_i } \right)\;^i \vec m_{PP} 
$.
The term $F_{SP}(m_i)$, for example, is proportional to $SP (m_i)$ along with the
overall acceptance and phase space for an SP interference term generated
in a given $m_i$ bin.  
Multiplying the ${}^i\vec D$ data vector by each $^i\vec m_\alpha$ produces a ``component" equation:
\[
\left( {\begin{array}{*{20}c}
   {^i\vec m_{SS}  \cdot {}^i\vec D }  \\
   {^i\vec m_{SP}  \cdot {}^i\vec D }  \\
   {^i\vec m_{PP}  \cdot {}^i\vec D }  \\
\end{array}} \right) = \left( {\begin{array}{*{20}c}
   {\vec m_{SS}  \cdot \vec m_{SS} } & {\vec m_{SS}  \cdot \vec m_{SP} } & {\vec m_{SS}  \cdot \vec m_{PP} }  \\
   {\vec m_{SP}  \cdot \vec m_{SS}  } & {\vec m_{SP}  \cdot \vec m_{SP}  } & {\vec m_{SP}  \cdot \vec m_{PP} }  \\
   {\vec m_{PP}  \cdot \vec m_{SS} } & {\vec m_{PP}  \cdot \vec m_{SP} } & {\vec m_{PP}  \cdot \vec m_{PP} }  \\
\end{array}} \right)\left( {\begin{array}{*{20}c}
   {F_{SS} \left( {m_i } \right)}  \\
   {F_{SP} \left( {m_i } \right)}  \\
   {F_{PP} \left( {m_i } \right)}  \\
\end{array}} \right)
\]
The formal solution to this is:
\[
\left( {\begin{array}{*{20}c}
   {F_{SS} \left( {m_i } \right)}  \\
   {F_{SP} \left( {m_i } \right)}  \\
   {F_{PP} \left( {m_i } \right)}  \\
\end{array}} \right) = \left( {\begin{array}{*{20}c}
   {\vec m_{SS}  \cdot \vec m_{SS} } & {\vec m_{SS}  \cdot \vec m_{SP} } & {\vec m_{SS}  \cdot \vec m_{PP} }  \\
   {\vec m_{SP}  \cdot \vec m_{SS}  } & {\vec m_{SP}  \cdot \vec m_{SP}  } & {\vec m_{SP}  \cdot \vec m_{PP} }  \\
   {\vec m_{PP}  \cdot \vec m_{SS} } & {\vec m_{PP}  \cdot \vec m_{SP} } & {\vec m_{PP}  \cdot \vec m_{PP} }  \\
\end{array}} \right)^{ - 1} \left( {\begin{array}{*{20}c}
   {^i\vec m_{SS}  \cdot \vec D_i }  \\
   {^i\vec m_{SP}  \cdot \vec D_i }  \\
   {^i\vec m_{PP}  \cdot \vec D_i }  \\
\end{array}} \right)
\]
This solution can be written as 
$F_{SS} \left( {m_i } \right) = {}^i\vec P_{SS}  \cdot \vec D_i $,
$F_{SP} \left( {m_i } \right) = {}^i\vec P_{SP}  \cdot \vec D_i $, and 
$F_{PP} \left( {m_i } \right) = {}^i\vec P_{PP}  \cdot \vec D_i $
, where the projection vectors are given by:

\begin{eqnarray}
\left( {\begin{array}{*{20}c}
   {\vec P_{SS} }  \\
   {\vec P_{SP} }  \\
   {\vec P_{PP} }  \\
\end{array}} \right) = \left( {\begin{array}{*{20}c}
   {\vec m_{SS}  \cdot \vec m_{SS} } & {\vec m_{SS}  \cdot \vec m_{SP} } & {\vec m_{SS}  \cdot \vec m_{PP} }  \\
   {\vec m_{SP}  \cdot \vec m_{SS}  } & {\vec m_{SP}  \cdot \vec m_{SP}  } & {\vec m_{SP}  \cdot \vec m_{PP} }  \\
   {\vec m_{PP}  \cdot \vec m_{SS} } & {\vec m_{PP}  \cdot \vec m_{SP} } & {\vec m_{PP}  \cdot \vec m_{PP} }  \\
\end{array}} \right)^{ - 1} \left( {\begin{array}{*{20}c}
   {^i\vec m_{SS} }  \\
   {^i\vec m_{SP} }  \\
   {^i\vec m_{PP} }  \\
\end{array}} \right)
\label{pvec}
\end{eqnarray}

We can modify the projector weights to take into account cut efficiency, acceptance
and phase space corrections  
required to convert say $F_{SP} (m_i )$ into $SP(m_i )$ as discussed in Reference \cite{helicity}.
This procedure scales the $^i\vec P_{SP}$ , for example, into a modified weight $^i\vec P'_{SP}$ 
which includes all acceptance, efficiency corrections and phase space effects.

The various projector dot products are implemented through a weighting technique. For example, 
if we are trying to extract the \SP{} interference in the $i^{th}$  \mkpi{} bin, we need to 
construct the dot product:

\[
{}^i\vec P'_{SP}  \cdot {}^i\vec D = \left[ {{}^i\vec P'_{SP} } \right]_1 {}^in_1  + \left[ {{}^i\vec P'_{{\rm{SP}}} } \right]_2 {}^in_2  +  \cdots \left[ {{}^i\vec P'_{SP} } \right]_{20} {}^in_{20} 
\]

We can do this by  making a weighted histogram of \mkpi{} where all events that are reconstructed in the first
$\cos{\theta}$ bin are weighted by ${}^i\left[\vec P'_{SP}\right]_1$ ; all events that are reconstructed in the second $\cos{\theta}$
bin are weighted by ${}^i\left[\vec P'_{SP}\right]_2$ etc. 

\mysection{\label{ambiguity} Amplitude ambiguity}

Table \ref{angular dist} shows the \costh{} dependences for each of the 3 direct and 3 interference
terms for the \ess{}, \pee{}, and \dee{} amplitudes.
\begin{center}
\begin{table}[htp]
\caption{The angular distributions for amplitudes up to and including D-wave. Here \costh{} is
the angle between the $K^+$ and $K^-$ in the \kpi{} rest frame. These terms are products
of the Wigner d-matrices \wigner{}.
\label{angular dist}}
\vskip .2in
\begin{tabular}{|c|c|c|c|}
\hline
SS & 1 & PP & $\cos^2{\theta}$\\
\hline
SP & $\cos{\theta}$ & PD & $\cos{\theta}~\left(3 \cos^2{\theta}-1 \right)/2$\\
\hline
SD & $\left(3 \cos^2{\theta}-1 \right)/2$ & DD & $\left(3 \cos^2{\theta}-1 \right)^2/4$ \\ 
\hline
\end{tabular}
\end{table}
\end{center}
The problem is that these six terms are not independent.  The relationship between them
is given in eq. (\ref{indy}).

\begin{eqnarray}
{{}^i\vec m_{SD}  = {{3~~{}^i\vec m_{PP}  - {}^i\vec m_{SS} } \over 2}}
\label{indy}
\end{eqnarray}

Hence one cannot make independent projectors of SS, PP, and SD, and must choose two out of 
these three or the inverse matrix of eq. (\ref{pvec}) will become singular.  Since it is 
known \cite{rlg} that the \kpi{} spectrum in \dpkkpi{} is dominated by the PP contribution (\krz{})
and SS contribution [\ksw{}], we made the choice of dropping the SD interference term and
choosing SS, PP, DD, SP, and PD. As a result, the five projectors we use can only partially block any potential SD term.
As long as there are no amplitude contributions beyond D-wave, one can show that a potential SD
interference contribution will contaminate both the SS and PP spectra but not the two interference
terms: SP or PD.

\begin{figure}[tbph!]
\begin{center}
\includegraphics[width=6.in]{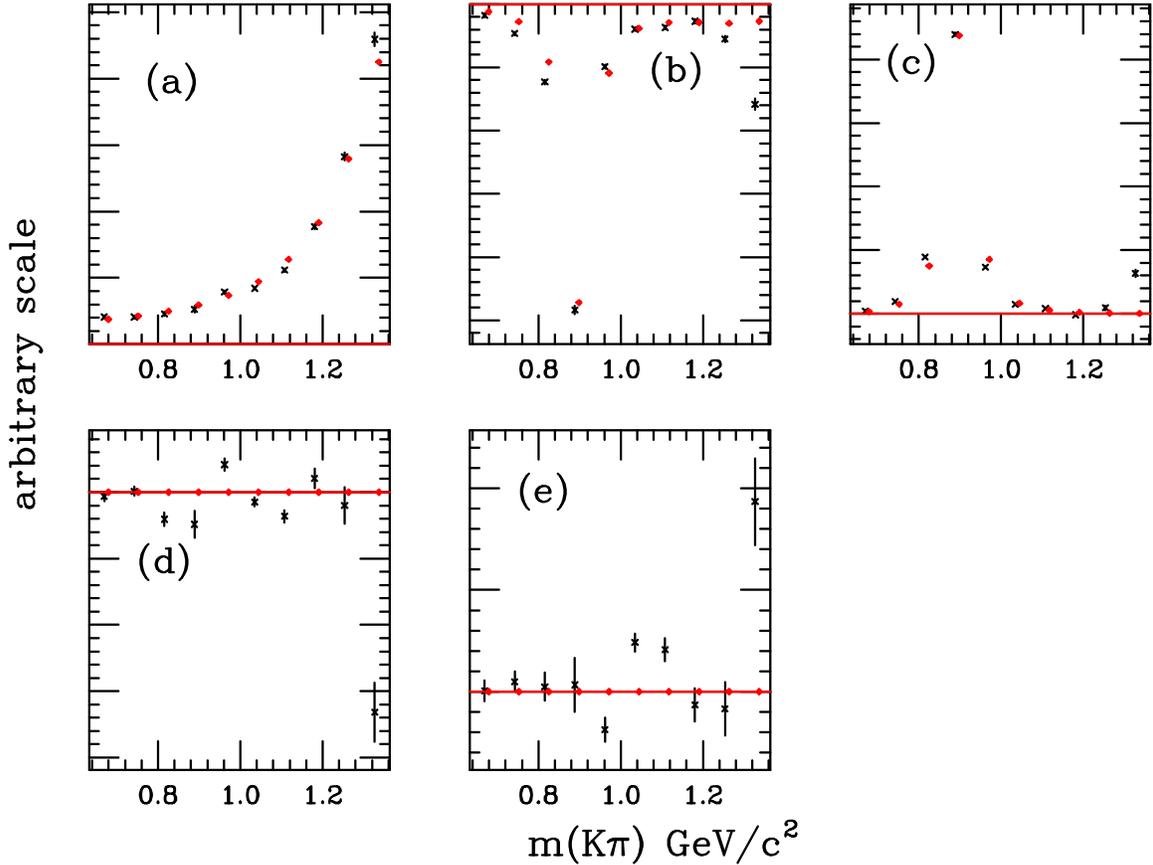}
\end{center}
\caption{The first points (``crosses") with error bars show the reconstructed spectra using the method described in Section \ref{proj}. The closely adjacent points (``diamonds") are the actual \mkpi{} spectra used in the simulation based on the model of Reference \cite{rlg}. These two Monte Carlo spectra are normalized near the peak bins of the prominent \krz{} present in the PP spectrum.
The plots are:
(a)~\Sint{} direct term,
(b)~\SP{} interference term,
(c)~\Pint{} direct term,
(d)~\PD{} interference term and
(e)~\Dint{} direct term.
\label{beforebias}}
\end{figure}

\mysection{\label{bias} The Bias Correction}
Figure \ref{beforebias} compares the five reconstructed spectra with the input spectra according to a simulation of the E687 model \cite{rlg}.  
The discrepancies between the reconstructed and input spectra are due to residual $\phi$ contamination in the tail beyond our 1050 \mevc{} anti-$\phi$ cut.   The discrepancies are most prominent at the high end of the \mkpi{} distribution (where the $\phi$ tail is the largest), but are relatively small on the scale of our statistical error elsewhere.

In order to correct for the $\phi$ tail that extends past \mkk{} = 1050 \mevc{}, we take the difference between the simulated input and reconstructed spectra as a ``bias" that we subtract from the data after normalizing the bias by the ratio of the \krz{} peak area of the PP spectra in the data to that in the simulation.\footnote{The normalization is actually done by finding a ratio that minimizes the $\chi^2$ between
the data and MC prediction at the peak and one bin on either side of the peak.} Varying the \kk{} amplitudes relative to the E687 model \cite{rlg} will change the bias correction. Uncertainty in these amplitudes is our major source of systematic error.

\mysection{\label{systematics} Systematic Errors}

The major source of systematic error in this analysis is due to possible uncertainty in the \kk{} system amplitudes. In Section \ref{bias} we discussed the method used to correct the amplitude spectrum in \kpi{} for residual \kk{} contributions past our anti-$\phi$ cut of $\mkk{} > 1050~\mevc{}$.  This method, however, depends  on our model for the \kk{} channel. The nominal results assume the $\phi$ amplitude measured in the E687 \cite{rlg} analysis, but we have considered variations in the result due to differences in the $\phi$ parameters as well as potential contributions from the \fz{}. Although the \dskkpi{} has an 11\% contribution from $\fz{} \pi^+$, there was no evidence for an $\fz{} \pi^+$ contribution in \dpkkpi{} in the E687 analysis.  

We have considered possible \fz{} contributions or $\phi$ parameter variants consistent with our data in four areas: (a) the fraction of \dpkkpi{} that appears as $\phi \pi^+$: (b) the shape of the \cosphidk{} distribution in the vicinity of the $\phi$: (c) agreement with the shape of the complete \mkk{} spectrum and (d) agreement with the observed populations in two dimensional bins of the \dpkkpi{} Dalitz plot.  Here \cosphidk{} is the cosine of the angle between the $K^+$ and the $\pi^+$ tracks in the $K^+ K^-$ rest frame.
In each case, the default model with a wide \ksw{} and no \fz{} amplitude was among the best models matching our data according to these tests. Essentially the same set of potential fit variants were selected by the four tests.

We found that a potential \fz{} when inserted with a phase of $\pi/2$ relative to the \krzb{}(890) satisfied our consistency criteria even with a fit fraction as large as 3.3\%. Such large \fz{} amplitudes would be more inconsistent with the data if they came in with a relative phase of zero. We construct the systematic error as the r.m.s. spread of all acceptable fit variants 
with \fz{} amplitudes up to 40\% of the \krzb{} amplitude that both matched the observed binned Dalitz plot populations to within 10$\sigma$ of the best model and matched the observed \cosphidk{} distribution in the vicinity of the $\phi$ with a confidence level exceeding $1 \times 10^{-4}$. Our default model matched the \cosphidk{} distribution with a confidence level of 18\%.  
Table \ref{results1} in Section \ref{results} compares the sizes of systematic
and statistical errors for our five amplitude products.  Generally the systematic error was found to be smaller than the statistical error. 

A variety of other checks of the results were made during the course of this work. 
We generated a Monte Carlo using the \mkpi{} amplitudes summarized in Table \ref{results1} and compared the
simulated to the observed  \costh{} distributions as a function of \mkpi{} as well as the 
simulated and observed \mkk{}, \mkpi{}, and \mnr{} global mass projections. Agreement was good. We also compared the fit results obtained by analyzing the data in different ways.
For example, we analyzed the data by constructing only three rather than five projectors and found consistent results with the \Sint{} , \Pint{}, and the \SP{} interference terms. We raised the anti-$\phi$ cut from $\mkk > 1050~\mevc{}$ to $\mkk > 1100~\mevc{}$ and found consistent results even though the errors in the high mass bins 
went up dramatically unless the D-wave projectors were excluded.

\mysection{\label{results} Results } 

\begin{figure}[htp]
\begin{center}
\includegraphics[width=6.in]{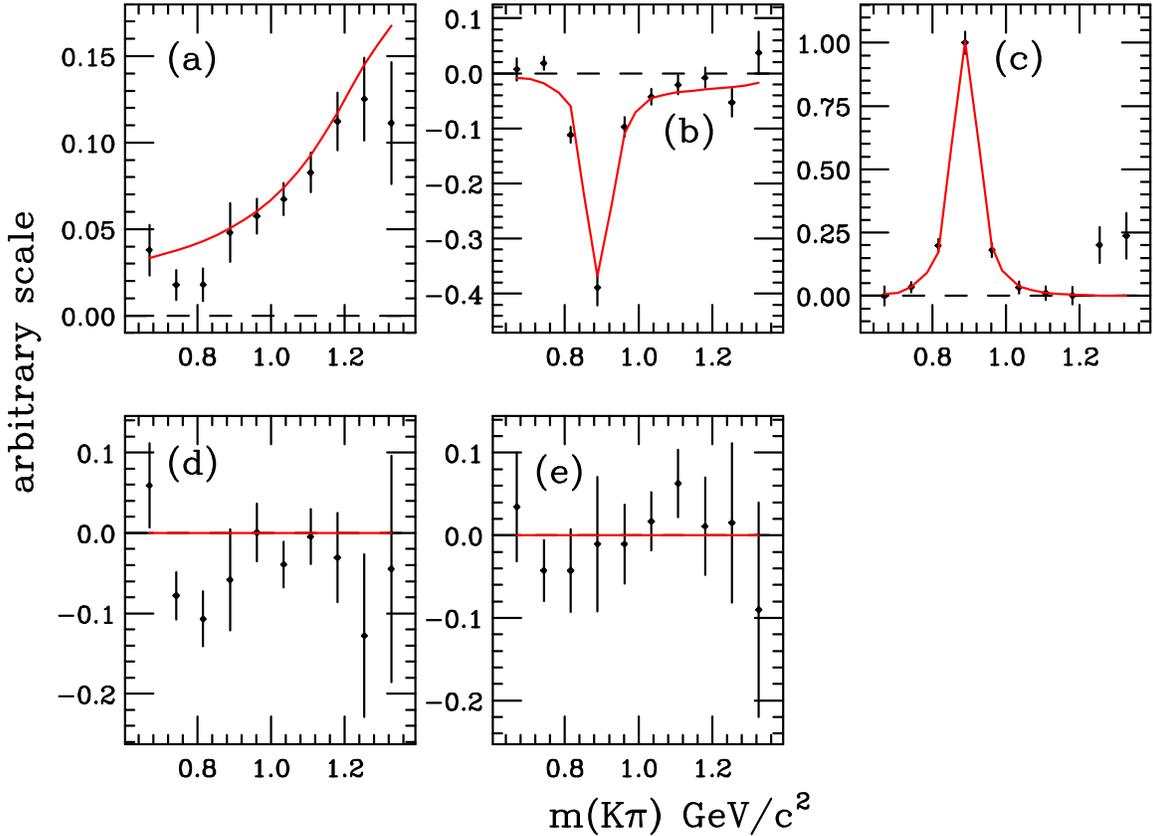}
\end{center}
\caption{ This figure shows the amplitudes measured in this analysis
including systematic errors in 73 \mevc{} bins. The default model described in the text are the curves.
The data has been normalized such that the \Pint{}
value at the \krz{} mass is 1 in plot (d).
The plots are:
(a)~\Sint{} direct term,
(b)~\SP{} interference term,
(c)~\Pint{} direct term,
(d)~\PD{} interference term and
(e)~\Dint{} direct term.
\label{sysall2}}
\end{figure}

Figure \ref{sysall2} and Table \ref{results1} show the results of this analysis.  
The relative correlation between the 5 amplitudes is typically less than $\pm 40\%$ except at the highest \mkpi{} bins
where they are as large as $\pm 65\%$. The comparison plot is based on the E687 analysis \cite{rlg} but with a much wider \ksw{} represented as a Breit Wigner resonance
with a pole at $m_0 = 1412~\mevc{}$ and a width of $\Gamma = 500~\mevc$ whereas the standard PDG~\cite{pdg} \ksw{} parameters
are $m_0 = 1414~\mevc{}$ and $\Gamma = 290~\mevc$. In order to roughly reproduce the results in figure \ref{sysall2} in plot (a), we also increased the magnitude of the \ksw{} amplitude by 40\% relative to that obtained from the E687 analysis. 
Hence the curves in fig. \ref{sysall2} are drawn using a model with a 41.9\% \ksw{} fit fraction compared to the $37 \pm 3.5 \%$ fit fraction quoted in Reference \cite{rlg}.
The use of a wider \ksw{} with a larger amplitude increased the level of the SP interference in the model by 45\% to a level in approximate agreement with our data as shown in plot(b). 

\begin{table}[htp]
\caption{Results for the five amplitude contributions. The first error bar is statistical and the second is systematic. All data are arbitrarily scaled by a common factor such that the peak \Pint{} value near the \krzb{} peak is set to 1.}
\vskip .15in
\begin{center}
{\scriptsize
\begin{tabular}{|l|c|c|c|}
\hline
\mkpi{}  & SS &  SP & PP  \\ 
(GeV/$c^2$) & & & \\
 \hline \hline 
0.670 & 0.038  $\pm$  0.015  $\pm$  0.002 & 0.007  $\pm$  0.020  $\pm$  0.002 & -0.001  $\pm$  0.037  $\pm$  0.004 \\ 
0.743 & 0.018  $\pm$  0.008  $\pm$  0.003 & 0.018  $\pm$  0.011  $\pm$  0.002 & 0.035  $\pm$  0.020  $\pm$  0.002 \\ 
0.816 & 0.018  $\pm$  0.009  $\pm$  0.003 & -0.112  $\pm$  0.014  $\pm$  0.005 & 0.198  $\pm$  0.025  $\pm$  0.007 \\ 
0.889 & 0.048  $\pm$  0.016  $\pm$  0.006 & -0.389  $\pm$  0.025  $\pm$  0.021 & 1.000  $\pm$  0.044  $\pm$  0.000 \\ 
0.962 & 0.058  $\pm$  0.009  $\pm$  0.004 & -0.097  $\pm$  0.014  $\pm$  0.011 & 0.181  $\pm$  0.024  $\pm$  0.011 \\ 
1.035 & 0.067  $\pm$  0.008  $\pm$  0.005 & -0.043  $\pm$  0.012  $\pm$  0.009 & 0.033  $\pm$  0.021  $\pm$  0.012 \\ 
1.108 & 0.083  $\pm$  0.009  $\pm$  0.007 & -0.021  $\pm$  0.012  $\pm$  0.013 & 0.011  $\pm$  0.022  $\pm$  0.015 \\ 
1.181 & 0.112  $\pm$  0.013  $\pm$  0.011 & -0.008  $\pm$  0.014  $\pm$  0.012 & 0.001  $\pm$  0.033  $\pm$  0.011 \\ 
1.254 & 0.125  $\pm$  0.021  $\pm$  0.011 & -0.053  $\pm$  0.023  $\pm$  0.011 & 0.201  $\pm$  0.067  $\pm$  0.023 \\ 
1.327 & 0.111  $\pm$  0.029  $\pm$  0.020 & 0.037  $\pm$  0.032  $\pm$  0.021 & 0.237  $\pm$  0.082  $\pm$  0.041 \\
\hline
\end{tabular}}
\vskip .02in
{\scriptsize
\begin{tabular}{|l|c|c|} 
\hline
\mkpi{} & PD &  DD  \\ 
(GeV/$c^2$) & &  \\
 \hline \hline 
0.670 & 0.059  $\pm$  0.053  $\pm$  0.003 & 0.034  $\pm$  0.066  $\pm$  0.007 \\ 
0.743 & -0.078  $\pm$  0.027  $\pm$  0.012 & -0.043  $\pm$  0.037  $\pm$  0.004 \\ 
0.816 & -0.107  $\pm$  0.034  $\pm$  0.004 & -0.043  $\pm$  0.046  $\pm$  0.021 \\ 
0.889 & -0.058  $\pm$  0.061  $\pm$  0.016 & -0.011  $\pm$  0.077 $\pm$  0.027 \\ 
0.962 & 0.001  $\pm$  0.031  $\pm$  0.017 & -0.011  $\pm$  0.041  $\pm$  0.025 \\ 
1.035 & -0.039  $\pm$  0.026  $\pm$  0.012 & 0.017  $\pm$  0.034  $\pm$  0.009 \\ 
1.108 & -0.004  $\pm$  0.033  $\pm$  0.009 & 0.063  $\pm$  0.040  $\pm$  0.007 \\ 
1.181 & -0.031  $\pm$  0.053  $\pm$  0.016 & 0.011  $\pm$  0.059  $\pm$  0.006 \\ 
1.254 & -0.128  $\pm$  0.099  $\pm$  0.022 & 0.015  $\pm$  0.095  $\pm$  0.018 \\ 
1.327 & -0.045  $\pm$  0.126  $\pm$  0.063 & -0.090  $\pm$  0.127  $\pm$  0.028 \\ 
\hline
\end{tabular}}
\end{center}
\label{results1}
\end{table}
\mysection{\label{summary} Summary and Discussion}
We presented a non-parametric amplitude analysis of the \kpi{} system in \dpkkpi{} decay using the technique described in Reference \cite{helicity}. There is no need to assume specific Breit-Wigner resonances, forms for mass dependent widths, hadronic form factors, or Zemach momentum factors. As described in Section \ref{proj}, a set of five weights were generated using Monte Carlo simulations that were designed to project the various amplitude contributions. The five amplitude contributions appear in just five weighted histograms in the \mkpi{} mass. Because this is an essentially one dimensional technique, we chose the \dpkkpi{} final state as a test case. According to an older, traditional Dalitz analysis done in E687 \cite{rlg} the \dpkkpi{} final state is particularly simple consisting of just $\phi \pi^+$, $\krz K^+$, and $\ksw K^+$.  Hence only one narrow resonance, the $\phi$, should contribute to the \kk{} channel that can be significantly reduced through an anti-$\phi$ cut such as $\mkk{} > 1050~\mevc{}$. This leaves an amplitude that depends primarily on the \kpi{} mass.  Another attractive feature of the \dpkkpi{} final state is that there should be a strong S-wave component in the \kk{} channel that E687 \cite{rlg} modeled as 
the \ksw{}.  Not that much is known about including broad S-wave resonances in charm Dalitz analyzes, and we indeed find considerable discrepancies between our non-parametric description of the  S-wave, \kpi{} amplitude in \dpkkpi{} and the standard, PDG \ksw{} parameterization used by E687 \cite{rlg}.

The other possibly significant difference in these results compared to the E687 analysis involves the ``glitch" in the first three bins of the \Sint{} spectrum. We have observed this effect in all variants of the fit we made in the course of understanding systematic errors. The same bins are also present as a glitch in the \PD{} interference term which is otherwise consistent with zero. If these are deemed to be significant, one explanation would be the presence of a small D-wave component at low \kpi{} masses which could account for both glitches through the mechanism discussed in Section \ref{ambiguity} while being small enough to escape notice in the direct \Dint{} terms where it is second order in the amplitude. Section \ref{ambiguity}  argues there is an ambiguity in the \costh{} distributions if one includes waves up to and including D-wave which cannot be eliminated by this or any other method. Hence an \SD{} interference term will also affect the \Sint{} spectrum.  

Although the \dpkkpi{} decay is an ideal case for an analysis of this kind, it might be possible to extend the analysis to the related \dskkpi{} decay\footnote{Care would need to be taken in eliminating or compensating
for the  $D^+ \rightarrow K^- \pi^+ \pi^+$ reflection shown in figure \ref{signal}, as
well as the \fz{} and $f_0(1710)$ contributions \cite{rlg}.} as well as $D^0 \rightarrow K^+ K^- \overline{K}^0$.  
The emphasis in $D^0 \rightarrow K^+ K^- \overline{K}^0$ decays would be on studying the \mkk{} spectrum after
applying cuts to minimize the effects of $K^\pm \bar K^0$ contributions such as the $a_o^\pm(980)$.
One could also use this technique to study the dipion amplitudes present in hadronic four-body decays such as 
$D_0 \rightarrow K^+ K^- \pi^+ \pi^-
\rightarrow \phi \pi^+ \pi^-$ decay.  In particular, one could compare the dipion spectra produced against longitudinally and transversely polarized $\phi$'s using techniques similar to those described in Reference \cite{helicity}.

\mysection{Acknowledgments}
We wish to acknowledge the assistance of the staffs of Fermi National
Accelerator Laboratory, the INFN of Italy, and the physics departments
of the collaborating institutions. This research was supported in part
by the U.~S.  National Science Foundation, the U.~S. Department of
Energy, the Italian Istituto Nazionale di Fisica Nucleare and
Ministero dell'Universit\`a e della Ricerca Scientifica e Tecnologica,
the Brazilian Conselho Nacional de Desenvolvimento Cient\'{\i}fico e
Tecnol\'ogico, CONACyT-M\'exico, the Korean Ministry of Education, and
the Korean Science and Engineering Foundation.

\end{document}